\documentclass[sigconf]{acmart}

\usepackage{booktabs}
\usepackage{multirow}
\usepackage{array}
\usepackage{diagbox}
\usepackage{tabularx}

\setcitestyle{numbers,sort&compress}

\AtBeginDocument{%
  }

\setcopyright{acmlicensed}
\copyrightyear{2017}
\acmYear{2017}
\acmDOI{}
\acmConference[Conference'17]{ACM Conference}{July 2017}{Washington, DC, USA}
\acmISBN{}


\title{Unified Heterogeneous Feature Interaction via Effective-Rank Scaling for Recommender Systems}

\author{Ao Zhang}
\affiliation{%
  \institution{Alibaba Group}
  \city{Hangzhou}
  \state{Zhejiang}
  \country{China}}
\email{aozhang@zju.edu.cn}

\author{Haochuan Li}
\affiliation{%
  \institution{Alibaba Group}
  \city{Hangzhou}
  \state{Zhejiang}
  \country{China}}
\email{lihaochuan.lhc@taobao.com}

\author{Boya Du}
\affiliation{%
  \institution{Alibaba Group}
  \city{Hangzhou}
  \state{Zhejiang}
  \country{China}}
\email{boya.dby@taobao.com}

\author{Jian Yang}
\affiliation{%
  \institution{Renmin University of China}
  \city{Beijing}
  \country{China}}
\email{yangjian0227@ruc.edu.cn}

\author{Qihang Fang}
\affiliation{%
  \institution{Institute of Automation, Chinese Academy of Sciences}
  \city{Beijing}
  \country{China}}
\email{fangqihang911@gmail.com}

\author{Yulin Xu}
\affiliation{%
  \institution{ Institute of Automation, Chinese Academy of Sciences}
  \city{Beijing}
  \country{China}}
\email{xuyulincs@126.com}

\begin{document}

\begin{abstract}
Large-scale industrial ranking models must jointly capture heterogeneous feature interactions, long-term user behaviors, and
multi-scenario dependencies under stringent online serving constraints. Existing architectures typically process sequential and
non-sequential features with separate modules, limiting bidirectional information exchange. Directly applying homogeneous self-attention to all semantic tokens can also be inefficient when sequential and non-sequential signals exhibit different effective-ranks (eRanks). Meanwhile, shared representations can overlook scenario-specific patterns and induce negative transfer between scenarios. In
this paper, we propose \textbf{HiRank}, a unified framework for multi-scenario recommendation that jointly models \textbf{H}eterogeneous feature \textbf{I}nteractions via effective-\textbf{Rank} scaling. HiRank integrates
gated sequential attention with heterogeneous feature interaction
within a unified backbone, enabling bidirectional fusion between
sequential and non-sequential representations. HiRank further uses
scenario-aware aggregation to compress heterogeneous low-order tokens into a compact representation. Distinct learned projections transform the full compact representation into scenario-aware
composite tokens with complementary global feature views and more
uniform information density, enabling more effective high-order
self-attention. Extensive experiments on industrial-scale datasets
demonstrate favorable performance scaling with parameter count
and consistent improvements over competitive baselines. Online
A/B tests further show a 2.27\% increase in orders per user.
\end{abstract}

\maketitle

\section{Introduction}

Ranking models~\cite{rankm2,covington2016youtube} in large-scale recommender systems are not limited to click-through rate (CTR) prediction. In production e-commerce systems, these models often need to jointly estimate multiple conversion-related targets, including purchase conversion rate (CVR) and add-to-cart rate. To estimate these targets, rankers integrate heterogeneous signals from user behavior histories and candidate item features. 

Prior research on recommendation ranking has largely progressed along two parallel lines: behavioral sequence modeling and feature interaction. The former focuses on extracting interest signals relevant to the candidate item from users' interaction histories. Representative advances include target-aware attention in DIN~\cite{zhou2018din}, GRU-based interest evolution in DIEN~\cite{zhou2019dien}, and efficient long-sequence compression in LONGER~\cite{chai2025longer}. The latter focuses on learning dependencies among high-dimensional features across multiple fields, including user attributes, item attributes, and contextual signals. Research in this direction has evolved from second-order interactions in factorization machines (FM)~\cite{rendle2010fm} and explicit higher-order feature crosses in DCNv2~\cite{wang2021dcnv2} to stacked FM blocks in Wukong~\cite{zhang2024wukong} and multi-head token mixing in RankMixer~\cite{zhu2025rankmixer}.
Despite progress in both directions, industrial ranking systems often connect these components through an \emph{encode-then-interact} pipeline. A sequence encoder first compresses a user's history into a fixed-dimensional representation, which a downstream interaction module then combines with non-sequential features. When this summary is the sole behavioral input to subsequent modules, they cannot directly access event-level sequence states to retrieve additional evidence within the same forward pass.

In parallel, the success of large foundation models~\cite{openai2023gpt4,rombach2022latentdiffusion} has motivated efforts to scale the capacity of recommendation models~\cite{guo2024scalingfrontiers,wang2025gpsd,zhang2024scalinglaw}. RankMixer~\cite{zhu2025rankmixer} and OneTrans~\cite{zhang2026onetrans} exemplify recent architectural advances in this direction. RankMixer combines efficient token mixing with token-specific transformations to scale heterogeneous feature interaction, while OneTrans unifies behavioral sequence modeling and feature interaction within a shared Transformer backbone. Both studies report favorable scaling behavior, with performance gains demonstrated through offline evaluations on production data and online A/B tests.
In CVR prediction, the relevance of behavioral signals depends on the candidate item, while their relative importance may vary across traffic sources and user segments. Given the sparsity of purchase conversion feedback, Taobao's homepage CVR model is trained on user-item samples with purchase feedback collected across the entire platform. This training setup motivates a scenario-aware CVR model that learns transferable patterns from platform-wide data while adapting to the characteristics of the target recommendation scenario.

In this paper, we propose \textbf{HiRank}, a unified framework
for multi-scenario recommendation that couples modeling of
the main user behavior sequence with long-term memory retrieval
and heterogeneous feature interaction. 
Inspired by the HSTU family~\cite{zhai2024hstu,ding2026ultrahstu}, HiRank maintains explicit main-sequence states for
gated sequential modeling, retains selected long-term behavioral
sequences as separate memories, and organizes non-sequential
features and pooled behavioral summaries into semantic tokens.
Its design considers representation diversity at two stages:
the variation of retrieval queries across users and the diversity
of composite tokens used for subsequent feature interaction.
For long-term memory retrieval, queries constructed from
relatively static profile features may concentrate in a limited
set of representation directions, restricting their effective
dimensionality across users. To preserve behavior-dependent
variation, HiRank initializes interest queries by attending to
the main sequence and uses the resulting
representations to retrieve evidence from the corresponding
long-term memories. A complementary candidate-conditioned
pathway reads the main sequence through shared gated-attention
projections and accesses the same memories from the candidate's
perspective. These pathways provide complementary behavioral
evidence for joint modeling with the main-sequence states and
semantic feature tokens.

HiRank integrates these representations through head-wise token
mixing and token-specific feed-forward transformations. When
multiple blocks are stacked, the interaction outputs update both
the main-sequence states and the queries used for subsequent
long-term retrieval. This feedback allows the evolving feature
context to guide further sequence modeling and memory access,
coupling behavioral evidence extraction with heterogeneous
feature interaction throughout the backbone.
To adapt the fused representations to different recommendation
scenarios, HiRank uses scene-aware queries to aggregate
the final token bank into a compact representation. Following
the composite-projection principle of
Hiformer~\cite{gui2023hiformer}, distinct learned projections
transform this complete representation into multiple composite
tokens for further attention-based interaction. Each composite
token has access to the full aggregated context while learning
a distinct projection of it. This construction aims to promote
complementary representation directions and a more balanced
singular-value spectrum across composite tokens, providing diverse
inputs for subsequent interaction. The refined shared
representation is then consumed by task-specific prediction
towers for multiple conversion-related objectives.

Our contributions are fourfold:
\textbf{(1) Unified heterogeneous feature interaction modeling.}
We propose HiRank, which integrates explicit main-sequence
states, query-based access to long-term memories,
and semantically grouped feature tokens within a shared
interaction backbone.
\textbf{(2) Main-sequence-guided long-term retrieval.}
We derive interest queries from the main sequence
to preserve behavior-dependent variation across users, and
complement them with candidate-conditioned queries for
long-term memory retrieval. When blocks are stacked, the
interaction outputs update both the main-sequence states
and retrieval queries, allowing the evolving feature context
to guide subsequent behavioral evidence extraction.
\textbf{(3) Scenario-aware high-rank latent interaction.}
We combine scene-aware aggregation with composite
projections of the complete aggregated representation to
construct multiple composite tokens. From an effective-rank
perspective, this design targets diverse representation
directions and a balanced spectrum for subsequent
attention-based interaction.
\textbf{(4) Serving and deployment.}
We conducted extensive experiments on Taobao's production dataset and deployed the proposed method in Taobao's recommendation system, which serves hundreds of millions of daily active users. The results demonstrate statistically significant increases in both the number of orders per user and gross merchandise value (GMV) per user.

\section{Related Work}

\subsection{Unified Sequence and Non-Sequence Modeling}

Unified ranking architectures differ in which representations remain explicit and how information flows between input types.
Feature-interaction architectures~\cite{gui2023hiformer,zhu2025rankmixer,ha2026unimixer} provide expressive operators for heterogeneous tokens.
Sequence-centric designs such as HSTU~\cite{zhai2024hstu} and ULTRA-HSTU~\cite{ding2026ultrahstu} emphasize event-level processing, with the latter exploring semi-local attention and separate transducers for different behavioral streams.
Joint modeling further requires an architecture that coordinates these representations throughout the backbone.

One approach preserves separate computational streams.
InterFormer~\cite{zeng2024interformer} maintains sequence and feature-interaction streams, exchanging selected summaries bidirectionally while retaining token-level states within each stream.
OneTrans~\cite{zhang2026onetrans} instead organizes both input types in a common causal token stream, sharing transformations across behavioral tokens and assigning token-specific parameters to heterogeneous feature tokens.
Its pyramid structure progressively reduces the sequential query positions processed by deeper layers.
Query-mediated architectures concentrate interaction in a compact set of evolving representations.
HyFormer~\cite{huang2026hyformer} retains behavioral sequences separately and alternates cross-attention decoding with global-query mixing.
MixFormer~\cite{huang2026mixformer} couples feature mixing with query-conditioned sequence aggregation, aligning feature heads with attention heads and constructing sequence key--value states through layer-specific feed-forward transformations.
These designs permit repeated access to behavioral evidence as the interaction context evolves.
HiRank adopts a hierarchical organization in which an evolving main-sequence state and source-specific long-term memories play distinct roles within the shared interaction backbone.

\subsection{Multi-Scenario Modeling}

Alongside backbone design, multi-scenario recommendation studies how a shared model adapts to heterogeneous recommendation contexts.
It is closely related to multi-domain learning, which balances shared patterns with domain-specific characteristics across traffic distributions.
Multi-task learning addresses a complementary dimension by sharing information across prediction objectives.
Foundational architectures include MMoE~\cite{ma2018mmoe}, which learns task-specific combinations of shared experts, and PLE~\cite{tang2020ple}, which progressively separates shared and task-specific representations to mitigate negative transfer.
For multi-domain CTR prediction, STAR~\cite{sheng2021star} combines shared central parameters with domain-specific parameters to capture both commonalities and distinctions across domains.

Subsequent methods refine adaptation and information sharing across domains and tasks.
PEPNet~\cite{chang2023pepnet} uses personalized priors to gate embeddings and hidden representations for multi-domain and multi-task adaptation.
M3oE~\cite{zhang2024m3oe} models common, domain-aspect, and task-aspect preferences through multiple expert modules and two-level fusion.
More recently, SAF~\cite{fei2026saf} combines sample-reweighted decorrelation with adaptive feature and expert gating to address spurious correlations in multi-domain multi-task recommendation.
The Multi-Distribution Learner (MDL)~\cite{mu2026mdl} represents features, scenarios, and tasks as tokens and lets scenario and task tokens query feature representations through domain-aware attention.
Combined with feature-token interaction and scenario--task fusion, this design supports joint multi-scenario and multi-task prediction.
HiRank pursues scenario adaptation through scenario-aware queries that select evidence from a shared bank of behavioral and non-sequential tokens.
The aggregated representation is then transformed into multiple composite tokens for interaction before task-specific prediction, connecting scenario-dependent evidence selection with the shared ranking backbone.

\begin{figure*}[t]
    \centering
    \includegraphics[width=0.90\textwidth,trim=55bp 39bp 144bp 96.9bp,clip]{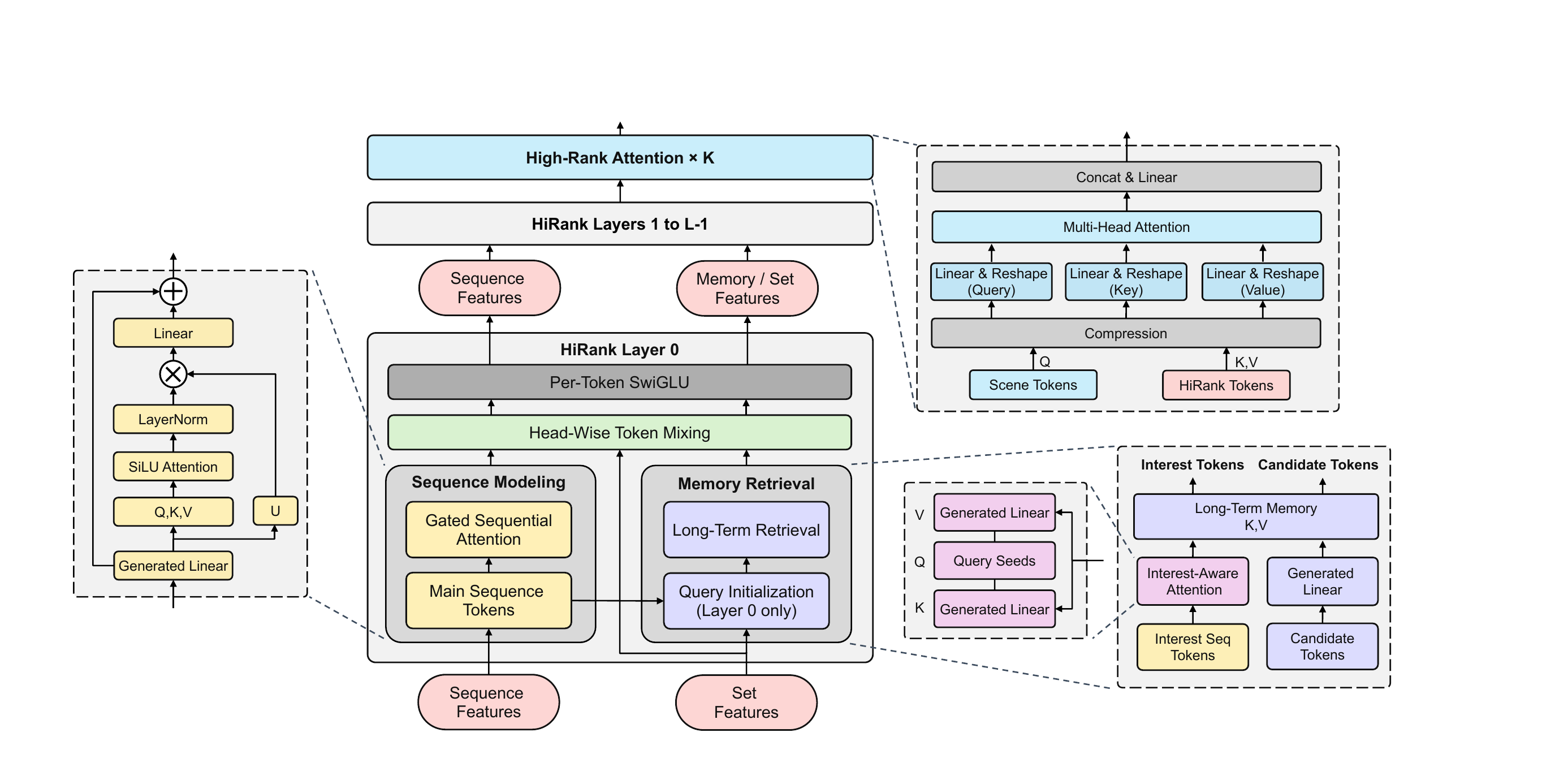}
    \caption{Overview of HiRank. Each block couples gated main-sequence modeling with long-term memory retrieval using main-sequence-guided and candidate-conditioned queries. Head-wise token mixing and per-token feed-forward transformations integrate the heterogeneous tokens. When blocks are stacked, the mixed states guide subsequent sequence modeling and memory retrieval. Scenario-aware queries aggregate the shared tokens into a compact representation, which is then projected into multiple composite tokens for multi-head attention.}
    \Description{HiRank model architecture. The center shows sequence features and set features entering repeated HiRank layers with sequence modeling, memory retrieval, head-wise token mixing, and per-token feed-forward networks. The left detail shows gated sequential attention with a gating branch and a residual connection. The lower-right details show query generation and main-sequence-guided and candidate-conditioned reads from long-term memory. The upper-right detail shows scene tokens querying HiRank tokens for compression, followed by query, key, and value projections, multi-head attention, and output projection.}
    \label{fig:architecture}
\end{figure*}

\section{Methodology}
\label{sec:methodology}

We consider multi-scenario ranking in an industrial e-commerce recommender system. For each recommendation request, the ranking model evaluates candidate items for a user and jointly predicts multiple conversion-related outcomes, including purchase conversion (CVR), add-to-cart, and other conversion-related objectives. We denote the corresponding task set by $\mathcal{T}=\{\mathrm{cvr},\mathrm{cart},\ldots\}$.

Given a user $u$ and a candidate item $i$, let $\mathcal{F}_{u,i}$ denote the non-sequential feature blocks describing the user, the candidate item, and the request context, and let $\mathcal{S}_u$ denote the user's historical behavior sequences. The prediction for task $t$ is formulated as
\begin{equation}
    \hat{y}_{u,i}^{\,t}
    = f_{\Theta}^{\,t}\!\left(\mathcal{F}_{u,i},\mathcal{S}_u\right),
    \qquad t\in\mathcal{T},
\end{equation}
where $\Theta$ denotes the trainable model parameters and $\hat{y}_{u,i}^{\,t}$ estimates the probability of a positive label $y_{u,i}^{t}=1$ for task $t$. Training samples are collected across multiple Taobao traffic sources, requiring the model to capture shared behavioral patterns while adapting to scenario-dependent feature relevance. Figure~\ref{fig:architecture} presents an overview of HiRank.

\subsection{HiRank Overview}
\label{sec:hirank-overview}

HiRank maintains a user main-sequence state, a candidate token, read-only long-term memories, and semantic feature tokens. Each block applies HSTU-inspired self-attention only to the main sequence; the candidate token reads the sequence keys and values through a separate cross-attention path with shared parameters. The updated states and long-term memory readouts are then integrated with feature tokens through head-wise mixing.
A Retrieval Query Initialization module constructs the interest and candidate-conditioned retrieval queries in the first HiRank layer. In subsequent layers, these queries are taken from the preceding layer’s retrieval outputs after token mixing and per-token feed-forward refinement. The mixed main-sequence and feature states are also propagated to the next layer.
The mixed sequence, retrieval, and feature slots form the next layer's initial states, allowing earlier feature interactions to guide subsequent sequence modeling and memory access.

After the final layer, the High-Rank Attention Block uses queries conditioned on scenario attributes to aggregate the shared token bank. It then projects the compact readout into multiple composite tokens, models their interactions, and feeds the resulting representation into task-specific towers. We also describe an optional user-item decoupled variant for sharing computation across candidates within a request.

\subsection{Features and Tokenization}
\label{sec:features-tokenization}

\subsubsection{Semantic feature tokens}
Non-sequential features and pooled behavior summaries are
organized into semantic groups, including user attributes,
item attributes, and request context.
Let $\mathcal{G}$ denote these groups. The corresponding feature tokens
are constructed as
\begin{equation}
    \mathcal{F}=\operatorname{Tokenize}(\mathcal{F}_{u,i},\mathcal{S}_u;\mathcal{G}).
\end{equation}
Within each group, the constituent feature blocks are
independently projected into $d$-dimensional tokens. Tokens from all
groups form $\mathcal{F}$.
We partition these tokens into user-side slots $\mathcal{F}_u$
and candidate-side slots $\mathcal{F}_i$.
These labels indicate the initial slot assignments; both sets
can contain cross-side information after joint mixing.
A separate projection of the candidate item features initializes
the candidate state $\mathbf{c}$.

\subsubsection{Main sequence and multiple long-term memories}
The main behavior sequence in $\mathcal{S}_u$ is projected event-wise into
$\mathcal{S}\in\mathbb{R}^{T\times d}$, preserving its
event-level representation for subsequent sequence modeling.
Each selected long-term history (e.g., click, add-to-cart, or purchase history)
is independently projected into a long-term memory
$\mathbf{M}_{j}\in\mathbb{R}^{T_j\times d}$,
with its corresponding valid-length mask.
The main-sequence state is updated across HiRank layers,
while the long-term memories retain their projected event states
throughout the forward pass.
We denote their collection by $\mathcal{M}=\{\mathbf{M}_j\}_{j=1}^{J}$.
Retrieval queries are initialized in the first HiRank layer,
as described next.

\subsection{HiRank Block}
\label{sec:hirank-layer}

\subsubsection{Gated Sequential Modeling}
Inspired by HSTU~\cite{zhai2024hstu,ding2026ultrahstu}, we combine gated self-attention over the user main sequence with candidate-to-sequence cross-attention.
Pointwise aggregated attention avoids the unit-sum constraint imposed by softmax. Multiplicative gating explicitly models interactions between local features and aggregated history.
The main-sequence projection is
\begin{equation}
    \mathbf{U},\mathbf{V},\mathbf{Q},\mathbf{K}
       =\operatorname{Split}\!\left(
          \operatorname{Linear}\!\left(\operatorname{LayerNorm}(\mathcal{S})\right)\right),
\end{equation}
where $\operatorname{Linear}$ is a position-shared affine projection and $\operatorname{Split}$ separates its four branches.
After splitting, SiLU is applied only to the gate branch; $\mathbf{U}$ in the gated update denotes this activated branch, while $\mathbf{Q},\mathbf{K},\mathbf{V}$ remain unactivated.
Writing the equations over valid positions and leaving head indices implicit, attention aggregates the projected values as

\begin{equation}
    \mathcal{S}'
       =\mathcal{S}+\operatorname{Linear}\!\left(
          \operatorname{LayerNorm}(\frac{\operatorname{SiLU}(\mathbf{Q}\mathbf{K}^{\top})\mathbf{V}}{N})\odot\mathbf{U}\right),
\end{equation}
here $N$ is the number of valid main-sequence positions.
After gated self-attention over the main sequence, the candidate token reuses the retained $\mathbf{K},\mathbf{V}$ projected from the block input $\mathcal{S}$.
Its own query and gate replace the sequence query and gate in the same computation, with the residual added to $\mathbf{c}$.
The candidate contributes no key or value to either attention read.
The input and output projections and both normalization stages are shared between the two paths within each block.
The updated states then enter joint token mixing with long-term memory readouts and feature tokens.

\subsubsection{Complementary Memory Retrieval}
In the first block, learnable query tokens $\mathbf{Q}_{\mathrm{seed}}$ attend to the main sequence $\mathcal{S}$ to produce interest queries $\mathbf{Q}_u$, partitioned into memory-specific groups $\mathbf{Q}_{u,j}$.
The updated candidate token initializes $\mathbf{Q}_{i,j}=\mathbf{c}'$ for each memory, stacked as $\mathbf{Q}_i$.
These two query sets retrieve complementary evidence from the same long-term memory $\mathbf{M}_j$:
\begin{equation}
\begin{split}
    \mathcal{R}_{u,j}
       &=\operatorname{CrossAttn}\!\left(
          \mathbf{Q}_{u,j},\mathbf{M}_j\right),\\
    \mathcal{R}_{i,j}
       &=\operatorname{CrossAttn}\!\left(
          \mathbf{Q}_{i,j},\mathbf{M}_j\right).
\end{split}
\end{equation}
Parameters are specific to each pathway and block.
After joint token mixing and per-token SwiGLU transformations, the updated representations from the two retrieval paths become the next block's queries.
The long-term memories retain their event representations throughout the stack.
This design separates persistent behavioral evidence from the evolving retrieval context, allowing later blocks to revisit the same histories in light of earlier feature interactions.

\subsubsection{Heterogeneous Token Interaction}
Grouping tokens by input type for notation, we write their concatenation along the token axis as
\begin{equation}
    \mathbf{X}
       =\operatorname{Concat}\!\left(
          \mathcal{S}',\mathcal{R},\mathcal{F},
          \mathbf{c}'\right).
\end{equation}
Here $\mathcal{R}$ stacks the outputs of both retrieval paths across long-term memories, and $\mathcal{F}$ contains all semantic feature tokens.
Following the mixing-and-reverting design of TokenMixer-Large~\cite{jiang2026tokenmixer}, $\operatorname{Mix}$ groups matching channel slices across tokens for head-wise interaction, while $\operatorname{Revert}$ restores the original token layout to preserve alignment in residual updates.
Independently parameterized head-wise SwiGLUs are followed by per-token SwiGLU transformations in the restored token layout:
\begin{equation}
\begin{split}
    \mathbf{X}'
       &=\mathbf{X}+\operatorname{Revert}\!\left(
          \mathbf{Z}+\operatorname{SwiGLU}_{\mathrm{head}}(\mathbf{\operatorname{Mix}\!\left(\mathbf{X}\right)})\right),\\
    \mathbf{X}^{(l)}
       &=\mathbf{X}'+
          \operatorname{SwiGLU}_{\mathrm{token}}\!\left(
             \operatorname{RMSNorm}(\mathbf{X}')\right).
\end{split}
\end{equation}
Here $\mathbf{Z}$ denotes the tokens obtained after mixing, $l$ indexes HiRank blocks, and $\mathbf{X}$ denotes the token-interaction input within the current block.
The first stage integrates behavioral and feature evidence across token positions within each channel group; the second recombines channels within each token using position-specific parameters.
When blocks are stacked, the updated representations corresponding to $\mathcal{R}$ become the next block's retrieval queries.

\subsection{High-Rank Attention}
\label{sec:high-rank-attention}
High-Rank Attention first forms a scenario-aware representation of $\mathbf{X}_{\mathrm{s}}$, then models interactions among composite tokens derived from this summary.

\subsubsection{Scenario-Aware Compression}
Inspired by Hiformer~\cite{gui2023hiformer}, we use composite projections to construct queries, keys, and values from the complete input representation for heterogeneous feature interaction.
Applying such projections to the concatenation of all backbone tokens requires large mapping matrices.
HiRank instead reduces the projection input dimension through scenario-aware compression.
We initialize scene tokens $\mathbf{Q}_s$ using the sample's scenario information and aggregate the final-layer output $\mathbf{X}^{(L)}$ through cross-attention:
\begin{equation}
    \mathbf{X}_{\mathrm{s}}
       =\operatorname{CrossAttn}\!\left(\mathbf{Q}_s,\mathbf{X}^{(L)}\right).
\end{equation}
The scene tokens serve as queries, while $\mathbf{X}^{(L)}$ provides keys and values.
Flattening $\mathbf{X}_{\mathrm{s}}$ gives the compact state $\mathcal{H}^{0}\in\mathbb{R}^{d_c}$ for subsequent composite-token interaction.

\subsubsection{Heterogeneous Attention}
Three independently parameterized composite projections~\cite{gui2023hiformer} transform the compressed state into queries, keys, and values.
Each composite token has distinct projection parameters and can combine all channels of the compressed state, allowing heterogeneous transformations of the same scenario-aware evidence.
For \(N_{\mathrm{t}}\) composite tokens, each projection matrix has size \(d_c\times(N_{\mathrm{t}}d_{\mathrm{t}})\), independent of the backbone token count.
At block $k=0,\ldots,K-1$, the projected tokens interact through multi-head attention:
\begin{equation}
\begin{split}
    \mathbf{Q},\mathbf{K},\mathbf{V}
       &=\operatorname{Reshape}\!\left(
          \operatorname{Split}\!\left(
             \operatorname{Linear}(\mathcal{H}^{k})\right)\right),\\
    \mathcal{H}^{k+1}
       &=\mathcal{H}^{k}+\operatorname{Linear}\!\left(
          \operatorname{Concat}\!\left(
             \operatorname{MHA}(\mathbf{Q},\mathbf{K},\mathbf{V})\right)\right).
\end{split}
\end{equation}
Attention models dependencies among these composite tokens, and the residual update integrates their interactions into the compact prediction state.

\subsubsection{Task Prediction}
Task-specific MLPs map the shared state to scalar logits, yielding $\hat{y}^t=\operatorname{Sigmoid}\!\left(\operatorname{MLP}_t(\mathcal{H}^{K})\right)$.
The training objective is
\begin{equation}
    \mathcal{L}
       =\sum_{t\in\mathcal{T}}\lambda_t\,
          \mathbb{E}\!\left[
             \operatorname{BCE}(y^t,\hat{y}^t)\right]
          +\mathcal{L}_{\mathrm{reg}}.
\end{equation}
Here $\lambda_t$ weights each task, and $\mathcal{L}_{\mathrm{reg}}$ regularizes model parameters.

\subsection{User-Item Decoupling}
\label{sec:ui-decoupling}

\begin{figure}[t]
    \centering
    \includegraphics[width=\columnwidth,trim=32bp 74bp 236bp 158bp,clip]{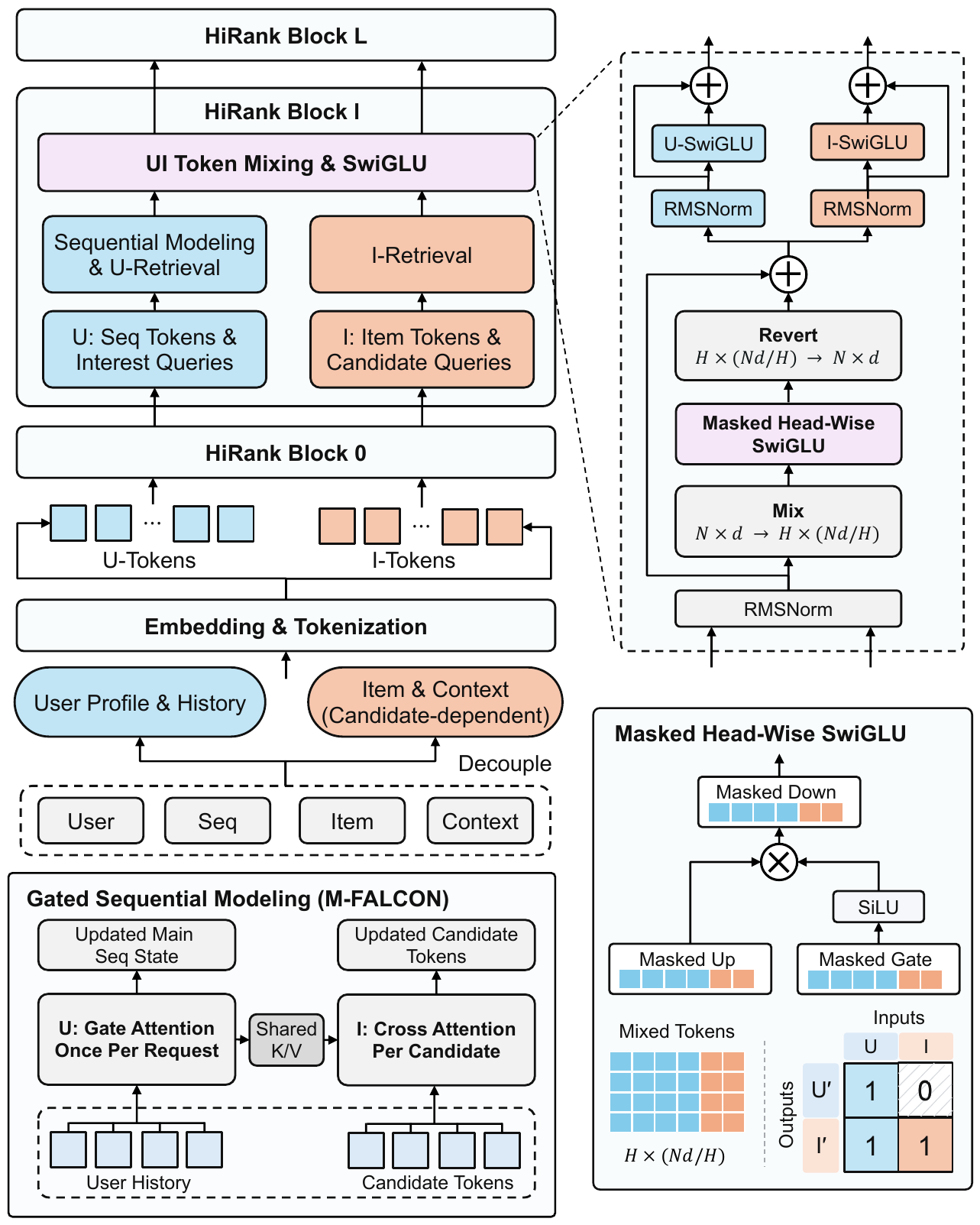}
    \caption{Optional user-item decoupled HiRank. User-side processing is shared within a request, while item-side processing remains candidate-specific. Blue and orange denote the user-side and item-side streams, respectively.}
    \Description{Blue user tokens and peach candidate tokens pass through embedding and tokenization and stacked HiRank blocks. Each block performs user-side sequential modeling and retrieval and candidate-side retrieval before UI token mixing and SwiGLU. The upper-right inset expands normalization, Mix, masked head-wise SwiGLU, Revert, and separate user/item token-local SwiGLU paths with residual connections. The lower-left M-FALCON inset shows shared user-history attention, per-candidate cross-attention, and shared K/V, producing updated main-sequence states and candidate tokens. The lower-right inset shows masked SwiGLU gating and an output projection, plus a dependency matrix with outputs as rows and inputs as columns: user outputs depend only on user inputs, while item outputs may depend on both sides.}
    \label{fig:ui-decoupling}
\end{figure}

Request-level batching (RLB)~\cite{guan2025makeitlong,guo2025roo} reduces redundant computation by sharing user-side processing across candidates within the same recommendation request.
Inspired by MixFormer~\cite{huang2026mixformer}, we formulate an optional variant that maintains candidate-independent user states while preserving user-to-item interaction, as illustrated in Figure~\ref{fig:ui-decoupling}.

\subsubsection{Token Decoupling}
For this variant, we partition backbone tokens into user-side and item-side streams according to whether their computation depends on the candidate item.
The user-side stream contains the main-sequence state, tokens from the interest-query retrieval path, and feature tokens shared across candidates within a request.
The item-side stream contains the candidate token, retrieval tokens conditioned on the candidate, and feature tokens encoding item attributes, user-item cross features, or context that varies across candidates.

\subsubsection{Directional Interaction}
Following the computation-sharing principle of M-FALCON~\cite{zhai2024hstu}, each layer computes main-sequence gated self-attention once per request and reuses its history K/V for candidate cross-attention.
Candidate queries read these keys and values without writing information back to the user sequence.

For token interaction, masked head-wise SwiGLU permits user-to-item information flow while blocking item-to-user feedback.
Its up, gate, and down projections use compatible user/item partitions to enforce the dependency pattern
\begin{equation}
    \mathbf{D}
       =\begin{pmatrix}1&0\\1&1\end{pmatrix},
\end{equation}
where rows index outputs and columns index inputs in $(u,i)$ order.

\addtocounter{table}{1}
\begin{table*}[!t]
\centering
\setlength{\tabcolsep}{3.5pt}
\setlength{\arrayrulewidth}{0.35pt}
\renewcommand{\arraystretch}{1.15}
\caption{Performance comparison. The best results for each performance metric are boldfaced, and the best baseline results are underlined. RI is computed relative to DCNv2 + DIN (Base) separately for Homepage and Main Search. Improv. denotes HiRank's improvement over the best baseline for each performance metric (reduction in $|\mathrm{PCOC}-1|$ for PCOC). Costs report dense parameters (M) and forward GFLOPs per candidate.}
\label{tab:offline-template}
\begin{tabularx}{\textwidth}{@{}>{\centering\arraybackslash}p{0.21\textwidth}|*{4}{>{\centering\arraybackslash}X}|*{4}{>{\centering\arraybackslash}X}|*{2}{>{\centering\arraybackslash}p{0.085\textwidth}}@{}}
\toprule
\multirow{2}[1]{=}{\diagbox[width=\linewidth,height=2.6\baselineskip,innerleftsep=3pt,innerrightsep=3pt,outerleftsep=0pt,outerrightsep=0pt,linewidth=0.35pt]{Method}{Metric}} & \multicolumn{4}{c|}{Homepage} & \multicolumn{4}{c|}{Main Search} & \multicolumn{2}{c}{Cost} \\
\cmidrule(l){2-11}
 & AUC $\uparrow$ & GAUC $\uparrow$ & PCOC & RI (\%) $\uparrow$ & AUC $\uparrow$ & GAUC $\uparrow$ & PCOC & RI (\%) $\uparrow$ & Params (M) & GFLOPs \\
\midrule
DCNv2 + DIN (Base) & 0.9136 & 0.7294 & 1.1866 & 0.00 & 0.8748 & 0.6764 & 0.9770 & 0.00 & 36.98 & 0.0428 \\
RankMixer~\cite{zhu2025rankmixer} + DIN & 0.9143 & 0.7473 & 1.1184 & 0.17 & 0.8755 & 0.6772 & 0.9674 & 0.19 & 350.15 & 1.002 \\
Hiformer~\cite{gui2023hiformer} + DIN & 0.9146 & 0.7518 & 1.0978 & 0.24 & 0.8754 & 0.6777 & 0.9829 & 0.16 & 392.97 & 1.243 \\
MixFormer~\cite{huang2026mixformer} & 0.9138 & 0.7488 & 1.1268 & 0.05 & 0.8752 & 0.6754 & 1.0214 & 0.11 & 341.80 & 1.161 \\
OneTrans~\cite{zhang2026onetrans} & \underline{0.9154} & 0.7592 & 1.1082 & \underline{0.44} & \underline{0.8759} & \underline{0.6778}  & 0.9896 & \underline{0.29} & 345.40 & 1.768 \\
HyFormer~\cite{huang2026hyformer} & 0.9148 & \underline{0.7615} & 1.0665 & 0.29 & 0.8758 &0.6775 & 0.9794 & 0.27 & 344.40 & 1.280 \\
\midrule
Base + PLE~\cite{tang2020ple} & 0.9138 & 0.7334 & \underline{1.0426} & 0.05 & 0.8754 & 0.6770 & \textbf{\underline{0.9912}} & 0.16 & 345.98 & 1.046 \\
MDL~\cite{mu2026mdl} & 0.9147 & 0.7439 & 1.0850 & 0.27 & 0.8756 & 0.6773 & 1.0684 & 0.21 & 343.62 & 0.990 \\
\textbf{HiRank} (ours) & \textbf{0.9163} & \textbf{0.7801} & \textbf{0.9865} & \textbf{0.65} & \textbf{0.8764} & \textbf{0.6787} & 0.9864 & \textbf{0.43} & 299.04 & 1.530 \\
\midrule
\textit{Improv.} & +0.0009 & +0.0186 & +0.0291 & +0.22 & +0.0005 & +0.0009 & -0.0048 & +0.13 & / & / \\
\bottomrule
\end{tabularx}
\end{table*}
\addtocounter{table}{-2}

The directional design preserves layer-wise user-to-item information flow, allowing candidate representations to incorporate shared user evidence throughout the backbone.
This extends computational reuse across both behavioral modeling and feature interaction, combining request-level efficiency with candidate-aware representation learning in a unified backbone.

\section{Experiments}
\label{sec:experiment-contract}

Through offline evaluations and online A/B tests, we aim to answer the following research questions (RQs).
\textbf{RQ1: Prediction performance.} Does HiRank improve CVR prediction over competitive baselines under comparable dense-parameter budgets?
\textbf{RQ2: Ablation and analysis.} How do HiRank's key components contribute to prediction quality and computational cost, and how does effective rank evolve across mixing blocks, memory retrieval, and High-Rank Attention?
\textbf{RQ3: Scaling validation.} How do width, token length, and depth in HiRank blocks and High-Rank Attention affect CVR prediction quality and forward computational cost?
\textbf{RQ4: Online A/B tests.} Can HiRank improve orders and GMV per user over the production ranker while satisfying serving-latency constraints?

\begin{table}[!htbp]
\centering
\caption{Approximate daily dataset statistics. B and M denote billions and millions, respectively. Interactions count purchase-positive samples and sparsity is their ratio to impressions.}
\label{tab:config}
\setlength{\tabcolsep}{2pt}
\renewcommand{\arraystretch}{1.15}
\begin{tabular}{c|c|c|c|c|c}
\toprule
Data & Impressions & Items & Users & Interactions & Sparsity \\
\midrule
Taobao & 2.4B & 87M & 206M & 94M & 3.92\% \\
Homepage & 438M & 17M & 91M & 4M & 0.92\% \\
\bottomrule
\end{tabular}
\end{table}
\addtocounter{table}{1}

\subsection{Experimental Setup}

\subsubsection{Dataset}
We construct an industrial-scale dataset for CVR prediction from user interaction logs collected across Taobao.
Table~\ref{tab:config} summarizes the dataset statistics.
Purchase conversion prediction is the main task, with cart addition prediction as an auxiliary task to provide additional supervision under sparse conversion feedback.
The dataset covers major traffic sources, including Taobao's main search and homepage recommendations, as well as dozens of smaller scenarios.
The production ranker is trained on pooled data from these scenarios to mitigate the sparsity of purchase feedback.

\subsubsection{Tasks and Metrics}
Offline evaluation focuses on CVR prediction, with cart addition providing auxiliary supervision during training.
We report global AUC, user-grouped AUC (GAUC), the Predicted Click Over Click (PCOC), and Relative Improvement (RI).
Global AUC measures discrimination across eligible test impressions.
GAUC measures within-user ranking quality by averaging per-user AUCs in proportion to each user's total evaluation weight.
PCOC closer to one indicates better agreement between predicted and observed conversions.
Following DIN~\cite{zhou2018din}, RI measures the improvement in global AUC relative to the baseline's margin over random guessing:
\begin{equation}
    \operatorname{RI}
       =\left(
          \frac{\operatorname{AUC}-0.5}
               {\operatorname{AUC}^{\mathrm{base}}-0.5}
          -1\right)\times100\%.
\end{equation}

\subsubsection{Baseline Methods}
We compare two baseline groups.
For ranking, our production baseline DCNv2 + DIN (Base) combines DCNv2~\cite{wang2021dcnv2} for feature interaction with DIN~\cite{zhou2018din} for target-aware history aggregation.
RankMixer + DIN and Hiformer + DIN replace the feature-interaction module with RankMixer~\cite{zhu2025rankmixer} and Hiformer~\cite{gui2023hiformer}, respectively.
We also include MixFormer, OneTrans, and HyFormer~\cite{huang2026mixformer,zhang2026onetrans,huang2026hyformer}.
Multi-task and multi-scenario baselines include Base + PLE and MDL~\cite{tang2020ple,mu2026mdl}.
Base + PLE replaces the prediction heads of the production baseline, while MDL is evaluated as a complete model.
Comparisons share features, histories, training examples, and tuning budgets.
The production ranker serves as an operational reference.

\begin{table}[t]
\centering
\setlength{\tabcolsep}{2.5pt}
\setlength{\arrayrulewidth}{0.35pt}
\renewcommand{\arraystretch}{1.15}
\caption{Ablation study of different HiRank components, where ``w/o'' is short for ``without''. The first two variants use DIN-style attention instead of the corresponding HiRank modules. The best results for each performance metric are marked in boldface.}
\label{tab:hirank-ablation}
\begin{tabularx}{\columnwidth}{@{}>{\centering\arraybackslash}p{0.60\columnwidth}|*{2}{>{\centering\arraybackslash}X}@{}}
\toprule
\multirow{2}[1]{=}{\diagbox[width=\linewidth,height=2.6\baselineskip,innerleftsep=3pt,innerrightsep=3pt,outerleftsep=0pt,outerrightsep=0pt,linewidth=0.35pt]{Method}{Metric}} & \multicolumn{2}{c}{CVR} \\
\cmidrule(l){2-3}
 & AUC & GAUC \\
\midrule
w/o Gated Attention &0.9160  &0.7623 \\
w/o Complementary Retrieval &0.9159  &0.7591 \\
w/o Token Interaction \& High-Rank Attention &0.9143  &0.7312 \\
w/o Token Interaction &0.9158  &0.7714 \\
w/o High-Rank Attention &0.9154  &0.7685 \\
\textbf{Full HiRank} & \textbf{0.9163} & \textbf{0.7801} \\
\bottomrule
\end{tabularx}
\end{table}

\subsubsection{Implementation Details}
All models use the same input features, behavior histories, and training data, with hyperparameters selected by validation CVR AUC under comparable dense-parameter budgets.
For HiRank, we set the token width to $d=256$, and use four long-term memories and 32 mixer heads.
The main behavior sequence is truncated to $T=48$ events.
The high-rank attention block uses a compressed dimension of $d_c=1536$ and $N_{\mathrm{t}}=24$ composite tokens of width $d_{\mathrm{t}}=256$.
Unless otherwise specified, the HiRank backbone uses two layers ($L=2$), and the high-rank attention stage uses one block ($K=1$).

\subsection{RQ1: Performance Evaluation}

Table~\ref{tab:offline-template} compares offline CVR performance and forward computational cost.
HiRank achieves the highest AUC and GAUC on both traffic sources.
The AUC and GAUC gains over the respective strongest baselines are 0.0009 and 0.0186 on Homepage, and 0.0005 and 0.0009 on Main Search, respectively.
HiRank also achieves the PCOC closest to one on Homepage.
The following ablation study examines the contributions of HiRank's key components.

Unlike language tokens represented in a shared vocabulary space, recommendation features are highly heterogeneous in both semantics and statistical properties.
Unified Transformer architectures such as OneTrans~\cite{zhang2026onetrans} demonstrate the potential of jointly modeling behavior sequences and feature interactions, yet their ablations also show that retaining all tokens at every layer can increase computation without consistent performance gains.
From a representational perspective, single-head attention aggregation satisfies $\operatorname{rank}(\mathbf{A}\mathbf{V})\leq\operatorname{rank}(\mathbf{V})$ and therefore cannot itself increase the algebraic rank of the value representations.
Moreover, pure softmax-attention stacks exhibit a bias toward token uniformity in the absence of residual connections and feed-forward transformations~\cite{dong2021attention}.
RankMixer~\cite{zhu2025rankmixer} offers a complementary mechanism through efficient cross-token channel regrouping and concatenation.
These considerations motivate a unified framework that combines attention with token mixing, balancing adaptive dependency modeling with efficient information exchange across heterogeneous features.

Industrial recommenders also maintain multiple long-term behavior histories.
Incorporating all such histories into the main sequence of a unified backbone can exceed practical computation and serving-latency budgets.
Alongside candidate-aware retrieval, we derive user-interest queries from the main behavior sequence to provide a complementary interest perspective.
This choice exploits the fine-grained behavioral context preserved in the main sequence, mitigating the information compression that may arise from aggregated profile statistics and providing a richer basis for interest retrieval.
Finally, High-Rank Attention can be viewed as a mechanism for scenario-specific representation learning, conceptually related to the sharing and specialization principles of PLE~\cite{tang2020ple}.
It generates multiple composite projections from a common scenario-aware compressed representation, allowing attention to operate on learned latent views that already integrate heterogeneous information.
Heterogeneous attention then models higher-order interactions among these views.
Our experimental results further support the effectiveness of this design.

\subsection{RQ2: Ablation and Analysis}

\begin{figure}[!t]
    \centering
    \includegraphics[width=\columnwidth]{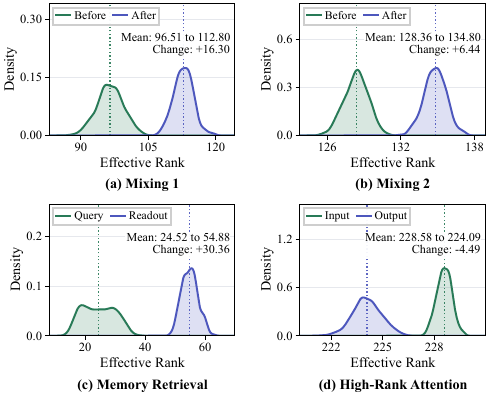}
    \caption{Pooled eRank over 354 aligned training steps: (a,b) mixing blocks; (c) memory retrieval; (d) High-Rank Attention. Dotted lines mark means.}
    \Description{Four density plots compare pooled effective rank before and after complete transformations. Mean eRank changes from 96.51 to 112.80 in the first mixing block, from 128.36 to 134.80 in the second, from 24.52 to 54.88 across user memory retrieval, and from 228.58 to 224.09 across the High-Rank block.}
    \label{fig:erank-analysis}
\end{figure}

\subsubsection{Ablation Study}
We perform an ablation study to quantify the contributions of HiRank's key components.
Table~\ref{tab:hirank-ablation} summarizes the following variants.
\textbf{Behavioral modeling variants:} i) replacing gated attention with DIN-style target attention over the main behavior sequence; ii) replacing complementary retrieval with DIN-style target attention over the same long-term histories, conditioned only on the candidate item.
\textbf{Interaction variants:} i) removing both token interaction (token mixing and per-token SwiGLU) and High-Rank Attention, and directly feeding the sequence, retrieval, and feature representations into the prediction MLP; ii) removing token interaction while retaining High-Rank Attention and the subsequent prediction modules; iii) retaining the complete HiRank blocks but removing High-Rank Attention, with the backbone outputs passed directly to the prediction MLP.
As shown in Table~\ref{tab:hirank-ablation}, full HiRank achieves the best AUC and GAUC among the evaluated variants.
Replacing gated attention or complementary retrieval with DIN-style attention reduces both metrics, supporting the value of modeling sequence dependencies and accessing complementary long-term evidence.
Removing token interaction or High-Rank Attention also degrades performance, with their joint removal yielding the largest AUC and GAUC drops of 0.0020 and 0.0489, respectively.
These results highlight the complementary roles of behavioral modeling and heterogeneous feature interaction in CVR prediction.

\subsubsection{Effective Rank Analysis}
Inspired by RankElastor~\cite{li2026rankelastor}, we characterize spectral changes using entropy effective rank~\cite{roy2007effective}.
For a nonzero, column-centered representation matrix $\mathbf A$, let $\sigma_j$ denote its positive singular values:
\begin{equation}
    \operatorname{eRank}(\mathbf A)
       =\exp\!\left(-\sum_j p_j\log p_j\right),
    \quad p_j=\frac{\sigma_j}{\sum_r\sigma_r}.
\end{equation}
Each logged tensor is flattened along its leading dimensions, independently subsampled to at most 256 rows, and column-centered.
To examine how HiRank integrates heterogeneous evidence while maintaining representation diversity, we analyze entropy-based effective rank at key stages.

Figure~\ref{fig:erank-analysis} reports pooled eRank distributions over aligned training steps.
The main-sequence representations have higher mean eRank after the two complete mixing blocks, increasing from 96.51 to 112.80 and from 128.36 to 134.80, respectively.
For user long-term memory retrieval, the retrieved representations also exhibit higher mean eRank than the input queries, rising from 24.52 to 54.88.
These shifts indicate greater spectral diversity in the corresponding representations under our pooled diagnostic.
For subsequent high-order interaction, independently parameterized dense projections transform the scenario-aware compact state into composite tokens, allowing each token to draw on all channels of the state.
Across the complete High-Rank Attention block, mean eRank changes from 228.58 to 224.09, a relative decrease of 1.96\%.
This modest reduction is consistent with largely retaining the compact state's spectral diversity during interaction refinement.
Together, these observations support HiRank's design of enriching representations through mixing and memory retrieval while limiting the reduction in spectral diversity during subsequent high-order interaction.

\subsection{RQ3: Scaling Validation}

\begin{figure}[!t]
    \centering
    \includegraphics[width=\columnwidth]{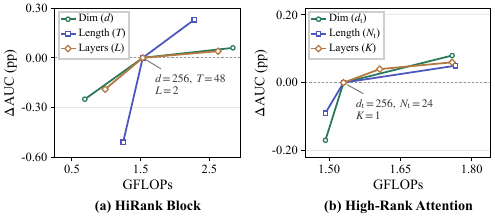}
    \caption{Scaling study of (a) HiRank Block and (b) High-Rank Attention.}
    \Description{Two side-by-side plots show scaling curves for dimension, token length, and layer count. The vertical axis is AUC difference in percentage points relative to default HiRank, whose AUC is 0.9163; negative values indicate lower AUC. The panels use different vertical scales. The horizontal axis is estimated full-model forward matrix GFLOPs per candidate. Every curve includes the default configuration, with two HiRank blocks and one High-Rank Attention block, at zero AUC difference. The largest gains are 0.23 percentage points from extending the main sequence to 96 events and 0.08 percentage points from increasing composite-token width to 1024.}
    \label{fig:scaling-validation}
\end{figure}

Figure~\ref{fig:scaling-validation} independently varies width, length, and depth for HiRank Block ($d,T,L$) and High-Rank Attention ($d_{\mathrm{t}},N_{\mathrm{t}},K$), fixing other settings.
We test $d\in\{128,256,384\}$, $T\in\{24,48,96\}$, $d_{\mathrm{t}}\in\{128,256,1024\}$, $N_{\mathrm{t}}\in\{12,24,96\}$, and $L,K\in\{1,2,4\}$.
The horizontal axis reports estimated full-model forward matrix GFLOPs per candidate.
The reference configuration uses $(d,T,L)=(256,48,2)$ and $(d_{\mathrm{t}},N_{\mathrm{t}},K)=(256,24,1)$ at 1.530 GFLOPs.
For HiRank Block, sequence length has the largest effect on AUC among the tested configurations ($T=96$: +0.23 pp; $T=24$: $-0.51$ pp), while increasing width or depth yields gains of at most +0.06 pp.
For High-Rank Attention, both composite-token width and token count contribute to performance.
At comparable costs, $d_{\mathrm{t}}=1024$ and $N_{\mathrm{t}}=96$ yield AUC gains of +0.08 and +0.05 pp, respectively, while $K=4$ yields +0.06 pp.

\subsection{RQ4: Online A/B Tests}
\label{sec:online-ab}

We conduct an online A/B test on Taobao Homepage traffic against the production ranker, using persistent user-level randomization while holding non-ranking policies fixed.
Orders/u is the primary outcome, and GMV/u uses the same assigned-user denominator and observation window.
Eligibility is determined before randomization, and both metrics include assigned users with no subsequent orders.
\begin{table}[t]
\centering
\setlength{\tabcolsep}{2.5pt}
\caption{Online A/B results: HiRank vs. the production baseline. Values are relative changes (\%). DAC denotes the number of daily unique purchasers. ${}^{*}$ denotes $p<0.05$ and ${}^{**}$ denotes $p<0.01$.}
\label{tab:hirank-online}
\begin{tabular}{l|cccc}
\toprule
Scenario & PV & Orders/u & GMV/u & DAC \\
\midrule
Homepage &$+0.45\% $&$+{2.27\%}^{**}$ &$+{2.86\%}^{*}$ & $+{2.13\%}^{**}$\\
\bottomrule
\end{tabular}
\end{table}
As shown in Table~\ref{tab:hirank-online}, HiRank increases PV, Orders/u, GMV/u, and DAC by 0.45\%, 2.27\%, 2.86\%, and 2.13\%, respectively, relative to the production baseline.
These gains indicate improved per-user transaction outcomes alongside an increase in the number of daily purchasing users.

We further observe a 0.85\% increase in the number of users clicking on Homepage recommendations and a 1.81\% increase in orders per 1,000 impressions.
These additional gains suggest that HiRank's benefits extend across user engagement and conversion efficiency.

\section{Conclusion}
We propose HiRank, a unified framework for multi-scenario recommendation that combines multi-behavior sequence modeling with heterogeneous feature interaction. Its backbone iteratively updates main-sequence states and complementary long-term retrieval queries, allowing feature context to guide subsequent behavior modeling. Scenario-aware aggregation and composite projections further construct complementary latent views for high-order interaction. Ablation studies support the complementary roles of the key components, while effective-rank diagnostics characterize representation diversity and scaling experiments show gains from longer sequences and increased model capacity. A user–item decoupled variant enables reuse of user-side computation across candidates within a request. 
Our method has been successfully
deployed on Taobao’s homepage feeds, where both offline evaluations and online A/B tests demonstrate improvements over the production baseline.

\label{www-main-end}

\end{document}